\documentclass[twocolumn,showpacs,preprintnumbers,amsmath,amssymb]{revtex4}

\usepackage{graphicx}
\usepackage{dcolumn}
\usepackage{bm}

\begin{document}

\preprint{Picasso/Discrimination}

\title{Discrimination of nuclear recoils from alpha particles with superheated liquids}

\author{F.~Aubin}
\altaffiliation[Present Address: ]{Department of physics, McGill University, Montr\'eal, H3A 2T8, Canada}
\author{M.~Auger}
\altaffiliation[Present Address: ]{Institut de physique, Universit\'e de Neuch\^atel, CH-2000, Neuch\^atel, Switzerland}
\author{M.-H.~Genest}
\altaffiliation[Present Address: ]{Fakult\"at f\"ur Physik, Ludwig-Maximilians-Universit\"at, D-85748 Garching, Germany}

\author{G.~Giroux}
\author{R.~Gornea}
\altaffiliation[Present Address: ]{Institut de physique, Universit\'e de Neuch\^atel, CH-2000, Neuch\^atel, Switzerland}
\author{R.~Faust}
\author{C.~Leroy}
\author{L.~Lessard}
\author{J.-P.~Martin}
\author{T. Morlat}
\altaffiliation[Present Address: ]{Centro de F\'isica Nuclear, Universidade de Lisboa, 1649-003, Portugal}
\author{M.-C.~Piro}
\author{N.~Starinski}

\author{V.~Zacek}
\altaffiliation[Contributing Author: ]{zacekv@lps.umontreal.ca}

\affiliation{D\'epartement de Physique, Universit\'e de Montr\'eal, Montr\'eal, H3C 3J7, Canada}

\author{B.~Beltran}
\author{C.B.~Krauss}
\affiliation{Department of Physics, University of Alberta, Edmonton, T6G 2G7, Canada}

\author{E.~Behnke}
\author{I.~Levine}
\author{T.~Shepherd}
\affiliation{Department of Physics \& Astronomy, Indiana University South Bend, South Bend, IN 46634, USA}

\author{P.~Nadeau}
\author{U.~Wichoski}
\affiliation{Department of Physics, Laurentian University, Sudbury, P3E 2C6, Canada}

\author{S.~Pospisil}
\author{I.~Stekl}
\author{J.~Sodomka}
\affiliation{Institute of Experimental and Applied Physics, Czech Technical University in Prague, Prague, Cz-12800, Czech Republic}

\author{K.~Clark}
\altaffiliation[Present Address: ]{Department of Physics, Case Western Reserve University, Cleveland OH 44106-7079}
\author{X.~Dai}
\author{A.~Davour}
\author{C.~Levy}
\author{A.J.~Noble}
\author{C.~Storey}
\affiliation{Department of Physics, Queens University, Kingston, K7L 3N6, Canada}

\date{\today}

\begin{abstract}
The PICASSO collaboration observed for the first time a significant difference 
between the acoustic signals induced by neutrons and alpha particles in a detector 
based on superheated liquids. This new discovery offers the possibility of improved background suppression and could be especially useful for dark matter experiments. This new effect may be attributed to the formation of multiple bubbles on alpha tracks, compared to single nucleations created by neutron induced recoils. 

\end{abstract}


\maketitle

\section{Introduction\protect}
The PICASSO dark matter experiment at SNOLAB uses the superheated droplet technique \cite{Bib1,Bib2}, which is based on the operation principle of the classic bubble chamber \cite{Bib3,Bib4,Bib5}. In this technique, the detector medium is a metastable superheated liquid and as described by Seitz \cite{Bib6}, the energy deposited by an ionizing particle causes heat spikes, which lead to the explosive formation of vapour bubbles on its track. In order to trigger bubble nucleation at a given temperature, a certain critical amount of energy has to be deposited within a certain critical length (which varies between tens of nm at high temperatures to a few $\mu$m at low temperatures).  Both quantities are functions of the surface tension and the superheat of the liquid, where superheat is defined as the difference between the vapour pressure of the liquid and the smaller external pressure. The larger the superheat, the more sensitive is the liquid to smaller energy depositions.

The amount of energy which is deposited within a critical length depends on the stopping power of the incident particle. Hence a distinction is possible between particles producing large ionization densities, such as nuclear recoils which are able to trigger bubble formation, and particles with reduced ionization density, such as cosmic muons,  $\gamma$- and $\beta$-rays, which are unable to do so. This background blindness is an important and unique feature of dark matter experiments using the superheated droplet technique, where very rare elastic scatterings of WIMPs (Weakly Interacting Massive Particles) produce highly ionizing nuclear recoils in the presence of less densely ionizing background events. The latter are more abundant in count rate by many orders of magnitude, but remain essentially undetectable due to this separation effect.             

The active detector material in PICASSO is superheated perfluorobutane, C$_{4}$F$_{10}$, which is sensitive to WIMP induced recoils at room temperature and at ambient pressure. By dispersing the superheated liquid in the form of droplets in polymerized gels or viscous liquids, these detectors can be kept continuously active. The burst of a droplet during a phase transition is accompanied by an audible click, which is recorded by piezoelectric transducers mounted at the detector wall. An added advantage of using fluorinated gases like C$_{4}$F$_{10}$, is their large $^{19}$F content. Due to its nuclear spin composition $^{19}$F is one of the most favorable nuclei for direct detection of WIMPs which undergo spin dependent interactions \cite{Bib7}. Threrefore  relatively small quantities of active target material already allowed competitive limits in this sector of dark matter searches to be obtained \cite{Bib5}.  Other dark matter experiments based on similar techniques are SIMPLE \cite{Bib8}, using droplets of C$_2$ClF$_5$ and CF$_3$I, and COUPP,  which operates a bubble chamber filled with CF$_3$I \cite{Bib9}.

Being insensitive to $\gamma$ and $\beta$ radiation is one of the assets of these detectors. However     
the superheated droplets are sensitive in the WIMP signal region to the higher specific 
energy losses of $\alpha$-particles and of nuclear recoils from neutron interactions. The $\alpha$-particles which are emitted by contaminations of the detector material with traces of U/Th and their radioactive daughter products constitute the most important background source. The background contribution from neutrons  is significantly reduced by operating the detector 2~km underground in the SNOLAB facility. In order to separate nuclear recoils (possibly from WIMP interactions) from alpha particle events, PICASSO has thus far exploited the fact that 
the temperature dependences of the interaction rates differ significantly in shape. Both contributions can then be fit to the recorded temperature spectrum and limits on a possible WIMP signal can be extracted. The sensitivity of this procedure is 
limited by the amount of $\alpha$-background and the precise knowledge of its temperature 
response. Therefore an event-by-event discrimination between $\alpha$-particle and WIMP induced 
recoil events would be very desirable and would add an important background suppression feature to this kind of dark matter 
search. 

Recent calibration runs using a new generation of PICASSO detector modules with droplet diameters between 50 and 280~$\mu$m,  showed that such a possibility exists as the acoustic signals contain information about the nature of the primary event. 
It was observed that in the acoustic frequency range corresponding to real signals the alpha signal is more intense than that of neutrons (and therefore of WIMPs). This suggests that the signal carries information about the first moments after bubble nucleation.  

The precise mecanism which causes this effect is at the moment not fully understood. A plausible
explanation could be that in the case of the more extended alpha tracks several proto-bubble nucleation sites contribute to the total signal, whereas the signals of the very localized nuclear recoils carry the imprint of only a single 
nucleation. That this effect was not noticed before might be related to the fact that earlier generations of 
PICASSO detectors worked with smaller droplet sizes with diameters of 10~$\mu$m or below and which therefore 
could sample only a fraction of the alpha range of about 35~$\mu$m. 

\section{Detector response\protect}
\begin{figure}
  \begin{minipage}[b]{8cm}
\begin{center}
  \includegraphics[width=8cm]{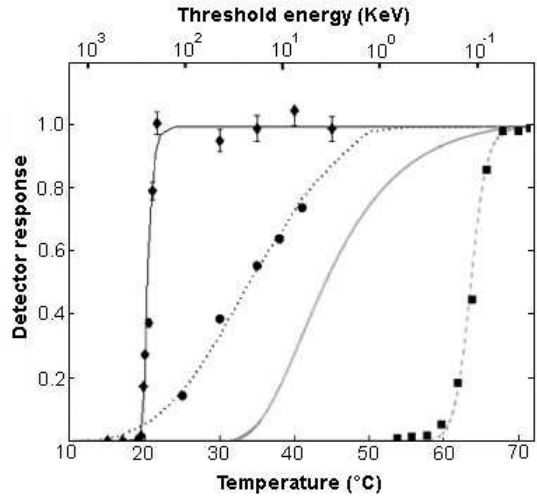}
   \end{center}
  \end{minipage}
\hspace{0.1cm}
 
\caption{Response to different kinds of particles as function of temperature for detectors loaded with C$_4$F$_{10}$ droplets of $\sim$200 $\mu$m in diameter. From left to right: alpha particles of 5.6~MeV in a detector spiked with $^{226}$Ra (data points fit with a continuous line); response to nuclear recoils from fast neutrons of an AmBe source (the points are data taken with det.93 and cover the error bars) compared to Monte Carlo simulations (dotted line); expected response to nuclear recoils following scattering of a 50 GeVc$^{-2}$ WIMP (gray continuous line); response to 1.275~MeV gamma rays of a $^{22}$Na source (dashed line). All responses are normalized to one at full detection efficiency. Temperatures are converted into threshold energies (upper x-axis) by using the n-test beam results from [10].}
\label{fig:resp}
\end{figure}

Droplet detectors are threshold counters and each individual superheated droplet 
acts as an independent bubble chamber. The threshold depends on temperature and pressure. 
At higher temperatures, the energy threshold for nuclear recoils is lower. The amount of energy deposited may be inferred by observing the threshold behaviour as the temperature is varied. The precise 
response to different forms of radiation was determined by calibrations with neutron sources, mono-energetic neutron beams, 
gamma ray sources and detector modules spiked with alpha emitters; the results are summarized in fig.~\ref{fig:resp} 
and described in detail in \cite{Bib10}. WIMP induced nuclear recoil energies are expected to be less 
than 100 keV and become detectable above 30$^{\circ}$C in PICASSO. Particles which produce low 
ionization densities, such as cosmic ray muons, $\gamma$- and $\beta$-rays only show up at temperatures above 
50$^{\circ}$C. They are well separated from the strongly ionizing neutron or WIMP induced recoils, 
which allows for an efficient suppression of such backgrounds \cite{Bib10}. 

The sensitivity to alpha particles that are contained entirely in the droplets starts 
with a sharp step at 21$^{\circ}$C. The $\alpha$-response curve shown in fig. 1 was obtained with a $^{226}$Ra spiked 
detector with large droplets where the emitters were distributed uniformly inside and outside 
of the droplets. The energy deposition necessary for nucleation at the threshold temperature of 21$^{\circ}$C 
is E$_{dep}$ = 252~keV, as determined in test beam measurements with mono-energetic neutrons \cite{Bib10}. 
At this temperature alpha particles  
become able to trigger a phase transition with their maximum dE/dx at the Bragg peak of 200~keV$\mu$m$^{-1}$ 
which gives an estimate for the critical length of L$_{c}$ = 1.3~$\mu$m.  
At higher temperatures the liquid becomes 
sensitive to smaller dE/dx on the alpha track and the recoiling daughter nucleus itself, but since 
the detector is already fully sensitive to alphas above threshold, the temperature 
response levels off to a plateau. 

\section{Measurements\protect}
The current generation of the PICASSO detectors consists of cylindrical modules of 14~cm diameter and 
40~cm height. The containers are fabricated from acrylic and are closed on top by stainless steel lids sealed 
with polyurethane O-rings. Each detector is filled with 4.5 litres of polymerized emulsion loaded with 
droplets of C$_4$F$_{10}$; the droplet volume distribution peaks at diameters around 200~$\mu$m. The active 
mass of each detector is typically around 85 g. The active part of each detector is topped by mineral 
oil, which is connected to a hydraulic manifold. After a measuring cycle of typically 30 hours, the detectors are recompressed 
at a pressure of 6 bar for about 15 hours in order to reduce bubbles to droplets and to prevent bubble growth which could 
damage the polymer. The operating temperature of the modules is uniform and controlled with a precision of $\pm$0.1$^{\circ}$C.   
Each detector is read out by nine piezo-electric transducers. The transducers are mounted at three 
different heights on the outside wall of each module, on a flat spot milled into the acrylic. The transducers are built with 
ceramic disks (Pz27 Ferroperm) 
with a diameter of 16~mm and 8.7~mm thickness. By comparing the signal arrival times of the different 
sensors, the position of each event can be reconstructed in 3D with a precision of several millimetres. 

A typical transducer signal starts with a fast rise, a maximum within the first 20-40~$\mu$s, followed by 
a series of slower oscillations, which settle down after several milliseconds. 
Studies in \cite{Bib11} of the dynamics of bubble explosions of individual superheated drops have shown 
that the first part of the pressure signal is related to the rapid growth of the vaporizing bubbles,
whereas a slow component is emitted when the entire droplet has evaporated into a large ringing bubble, 
performing damped oscillations. If a significant fraction of the acoustic signal is emitted during bubble 
growth shortly after nucleation, then the observed transducer signal should carry an imprint of the nature 
of the ionizing event. Later the primary bubbles will merge and vaporize the entire droplet into one single 
bubble and the original history of the event will be lost. 

\begin{figure}
\begin{minipage}[b]{8cm}
\begin{center}
     \includegraphics[width=8cm]{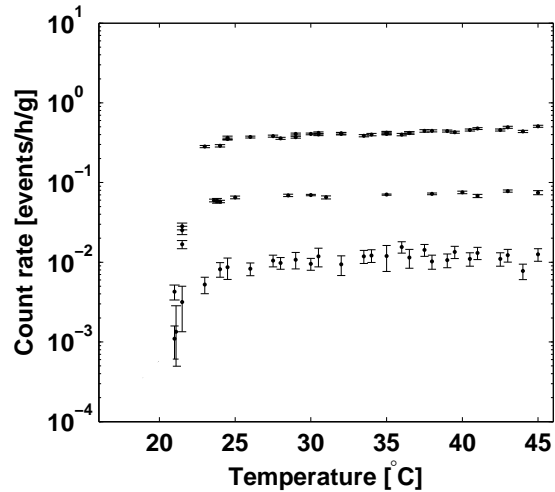}  
\end{center}
  \end{minipage}
 \caption{Background count rates as a function of temperature for detectors 76 (top), 93 (middle) and 71 (bottom) used in this analysis. The count rates, normalized by the active mass of each detector indicate the different levels of $\alpha$-background. The data were taken in the PICASSO installation at SNOLAB.}
\label{fig:curves}
\end{figure}

Evidence for such an effect was found in a comparison between calibrations with 
fast neutrons from an AmBe neutron source and alpha background data in the latest PICASSO set up at SNOLAB. 
PICASSO is currently in a development phase; with new detectors being installed underground, the electronics being tuned and optimized, and calibrations being performed. The figure of merit in this analysis is the amplitude of the signals recorded by the piezo-electric transducers. This is determined in part by the proximity of the transducer to the bubble site. To minimize solid angle effects of distributed events, only detectors with nine active transducers were used in this analysis. Although many detectors (16) were taking data, 
only a subset of 5 detectors (det. 71, 72, 75, 76, 93) had either enough active transducers or enough channels with well matched gains to be suitable for this study (during WIMP runs only 5 active channels are required). The background count rates of three of these detectors are shown
in fig.~\ref{fig:curves}. Their temperature profiles are similar to the response of the $^{226}$Ra spiked detector shown in fig. 1 and therefore are attributable to alpha particles. The thresholds are slightly shifted to higher temperatures, which is due to the increased pressure of 1.2 bar at the underground site. The natural neutron flux at SNOLAB is far below the current sensitivity of the detectors, which strengthens further the hypothesis that the recorded background events are due to $\alpha$-particles. 

A frequency filter with a cut-off below 18~kHz was applied to the signals 
and the maximum peak in each waveform of the nine transducers was determined. In the standard PICASSO analyses, 18~kHz is a typical value used in the bubble identification algorithms. It has been selected as a compromise between signal selection and electronic noise suppression (these filtered signals are smaller by a factor $\sim$ 3.5)   
\cite{Bib12, Bib13}.   

\begin{figure}
\includegraphics[width=8cm]{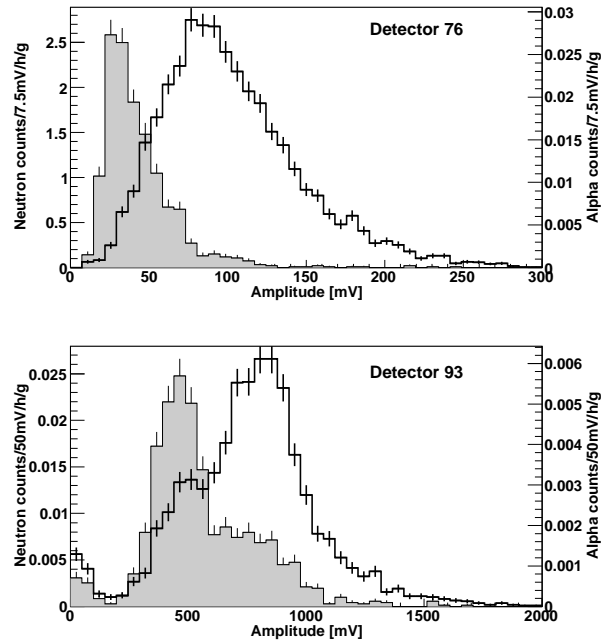}
\caption{\label{fig:disc3}Amplitude distributions of neutron and alpha particle induced events for detectors 76 (top) and 93 (bottom) with C$_4$F$_{10}$ droplets of $\sim$200 $\mu$m in diameter at 35$^{\circ}$C.  Plotted is the average of the peak amplitudes of nine transducers per detector. Count rates are given per active mass. A frequency filter with a cut-off below 18~kHz was applied. Neutron data are represented by shaded and $\alpha$-data by unshaded histograms. Det.~76 has a high $\alpha$-background and the AmBe neutron source was at 10~cm distance. The neutron calibration with det. 93 was performed with the source 1.5 m away; here the $\alpha$-background appears also in the calibration run as a shoulder at higher amplitude.}
\end{figure}

The distributions of the mean of the maximum amplitude of transducers are shown in fig.~\ref{fig:disc3} for detectors 76 and 93, both for  35$^{\circ}$C. In both figures the neutron events are concentrated at lower amplitudes (shaded histograms) and overlap partially with events taken during background runs (open histograms).  
As shown in fig.~2, detector 76 had a particularly high alpha background due to a fabrication accident (as did detector 75 with identical behavior, but not shown here). The neutron source was close to this detector (10 cm) and the neutron peak is well defined. Detector 93 has a reduced alpha background, but the neutron source in this measurement was located 1.5~m away, such that the neutron count rate during calibration was only about a factor two larger than the alpha rate. In this case the alpha component is also present during the neutron run and appears as a shoulder at higher amplitudes. The preamplifier gains were a factor of ten different in both measurements (x350 vs. x3500) resulting in sensitivities to pressure amplitudes of about 100~mV/Pa for det.~76 and 1000~mV/Pa for det.~93, respectively.

\begin{figure*}
\includegraphics[width=0.8\textwidth]{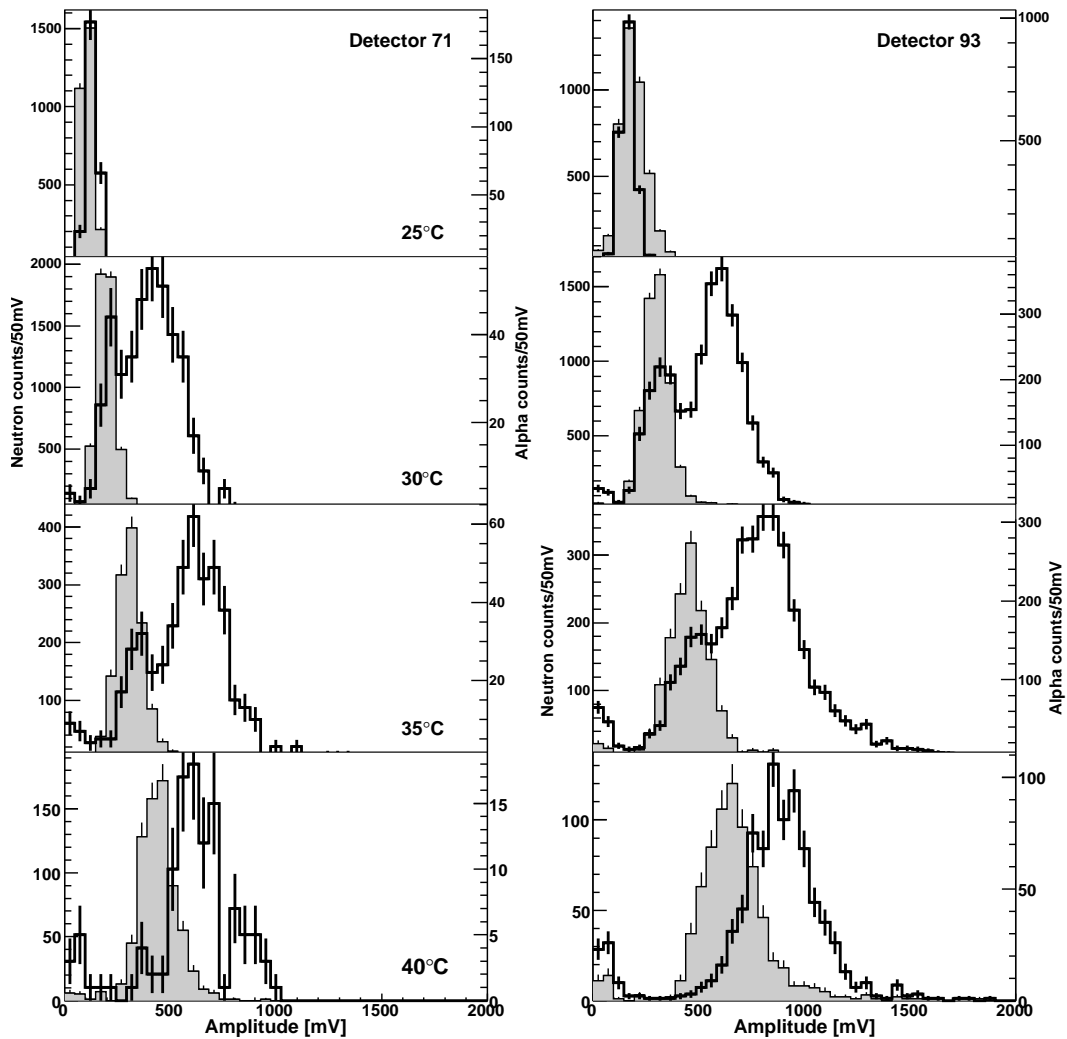}
\caption{\label{fig:disc4}Amplitude distributions of neutrons (shaded histograms) and alpha particle induced events (unfilled histograms) for detectors 93 (left) and 71 (right)  for 25$^{\circ}$C 30$^{\circ}$C, 35$^{\circ}$C and 40$^{\circ}$C. The analysis was performed as described for the data in fig.~\ref{fig:disc3}. The neutron data were obtained in a calibration run with an AmBe source at 10~cm distance to the detectors. To reduce the influence of gain instabilities the background runs have been taken immediately before and after the calibration at the respective temperature. A signal amplitude of 1000 mV corresponds to a pressure amplitude of 1 Pa.}
\end{figure*}

The time interval between calibration and background runs for a given temperature typically amounts to several weeks. During this time
gain instabilities or variations in acoustic couplings could occur leading to shifts of the amplitude distributions. In order to rule out such effects a special neutron calibration was performed with detectors 71, 72 and 93 where the alpha background data were taken immediately before and after the neutron source runs. The data were analysed as above described and the results are shown in fig.~\ref{fig:disc4} for two of the detectors and several temperatures (det.~72 is not shown because of poor $\alpha$-statistics). 

The peak of the neutron amplitude 
distribution is remarkably well defined for det.~71 and has a resolution at full width half maximum of about 20\%. With increasing temperature 
the recoil peaks move towards higher amplitudes at a pace of about 8\% per degree Celsius. 
At 25$^{\circ}$C the alpha and neutron responses seem to merge. But at
higher temperatures, such as 30$^{\circ}$C, 35$^{\circ}$C and 40$^{\circ}$C we observe for all detectors 
under investigation a clear change of the alpha response with respect to the neutron signal, similar to the effect shown in fig.~\ref{fig:disc3}. 
The alpha component appears to be split into peaks, with  a first peak coinciding with the 
neutron response and a second, larger peak, which is separated from the first one and 
is shifted towards higher amplitudes.  At higher temperatures the contribution of the first 
component tends to decrease, while the second increases. As in fig.~\ref{fig:disc3}, det.~93 shows also in these data a small shoulder to the right
of the neutron peak, which is compatible with the actual alpha background of this detectors.   

Although the data in fig.~\ref{fig:disc4} suffer from low statistics due to the amount of data currently available for analysis, particularly at 40$^{\circ}$C, it is interesting to derive an estimate of the alpha background rejection, which can be obtained by cutting the background at e.g. one full width at half maximum of the neutron peak: in order of increasing temperatures the alpha background can be rejected at a level of 46$\pm$5\%, 82$\pm$6\%, 81$\pm$6\% and 76$\pm$11\% for det.~71 and 4$\pm$0.4\%, 75$\pm$2\%, 76$\pm$2\% and 70$\pm$4\% for det. 93. 

Both the neutron and the alpha distributions have significant numbers of outliers, in particular near zero amplitude in fig.~\ref{fig:disc4}. These signals correspond to good, particle induced events in the unfiltered sample, but have waveforms with very large amplitudes, which saturate the preamplifiers over most of the signal duration (see discussion in sect. IV).  

Since alpha emitters are uniformly distributed throughout the detector volume, whereas events triggered by 
neutrons are more concentrated in the detector region close to the source due to attenuation and solid angle 
effects, we had to verify that the averaging process over the nine transducers compensates for these effects. 
For this we used the localization capability of our detector modules to select a uniformly distributed neutron-induced 
event distribution in the detector volume. A comparison of this with the non-uniform event sample showed that the 
discrimination effect remained unchanged. 
	
Another difference between neutron and background runs is the higher count rate in the detectors during neutron runs. 
Since more than one thousand bubbles are created in a calibration run, sound attenuation effects
due to the increasing number of bubbles in the detector volume might lead to decreasing signal amplitudes with time. 
This could bias the mean amplitude distribution in a neutron run towards lower amplitudes, thus imitating a separation 
effect. In order to investigate this aspect we traced the time dependence of the mean amplitudes of the detectors used in 
this study, but found no significant change in time.        

\section{Discussion\protect}

Above 21$^{\circ}$C fully contained alpha tracks in the droplets should develop one nucleation centre in the region of the Bragg peak; above 25$^{\circ}$C two vaporisation centres should appear: at the beginning of the track due to the strongly ionizing recoil nucleus and at the end of the track due to the Bragg peak. If the alpha emitters are located outside of the droplets, in the detector matrix, then the recoil nuclei would not contribute. However in both cases, fast alpha particle would be able to start nucleation above 25$^{\circ}$C and with increasing temperature the probability of multiple bubble formation on the ionizing alpha track should further increase due to the repeated occurrence of energy depositions, which meet the necessary nucleation criteria. Monte Carlo simulations do predict this. 

Assuming that the density of primary bubbles is proportional to the number of critical lengths covering the entire alpha track, the nucleation model of Seitz \cite{Bib6} predicts a roughly exponential increase of the bubble density with increasing temperature, up to temperatures where the liquid becomes sensitive to minimum ionizing particles (MIPs). An increase of the bubble density with temperature was also observed in bubble chambers (although investigated only in a limited temperature range) \cite{Bib14}. It is also known from bubble chamber operation that minimum ionizing particles can produce bubble densities from 20 to 200 bubbles per cm and more \cite{Bib15}. Moreover it was found that the bubble density scales with the stopping power of the incident particles \cite{Bib16}. 

Therefore, if one assumes for example 20 bubbles per cm for the bubble density of minimum ionizing particles in C$_4$F$_{10}$, a 5 MeV alpha particle would produce a density of around 0.6 nucleations per $\mu$m, i.e. 21 primary bubbles per alpha track above 60$^{\circ}$C in the MIP senitive region (fig.~1). Using the fact that a fast alpha particle is capable to produce one nucleation around 25$^{\circ}$C, an exponential interpolation between the two temperature regimes would predict approximately 2.3 nucleations at 35$^{\circ}$C and 3.6 nucleations at 40$^{\circ}$C on a track of 35 $\mu$m length. These bubbles would appear in addition to the nucleation at the Bragg peak, if the alpha particles enter the droplet or in addition to the Bragg peak and recoil induced bubbles, if the alpha track is contained.  

According to the data we see a trend, however not a good agreement with the proposed model. At e.g. 35$^{\circ}$C the data of detectors 71 and 93 indicate an alpha signal in the single-nucleation region coinciding with the neutron response and a second, wider distribution centred at about twice the mean amplitude of the neutron recoils. The component of the alpha signal in the neutron region might be due to a finite probability to produce single bubbles on the alpha track at this temperature or due to entering or leaving tracks, which are too short in the droplet to give rise to more than one nucleation.  However at higher temperatures the data do not show the expected continuous increase in bubble multiplicity. One reason might be, that the assumed exponential scaling of the bubble density with temperature is not valid. Also there might exist a mechanism, which quenches the formation of multi-bubble sound emission. 

Another explanation could be that electronic saturation due to the increasing amplitudes of the waveforms affects the discrimination power. This effect was studied by forcing raw waveforms to saturate and the filtered amplitude distribution was recorded as a function of the cut-off level. It was found that this effect becomes important above 40$^{\circ}$C, where also the number of filtered outlier events at very low amplitudes starts to increase strongly. The fact that the alpha events of det.~76 (with reduced gain) reach out to larger amplitudes, might be an indication in support of this hypothesis. In the future it will be interesting to explore higher temperatures in systematic studies with properly adjusted preamplifier gain.        

Another open question is the choice of the optimum lower frequency cut-off. By varying the cut-off frequency in the range from 6 kHz to 24 kHz preliminary studies showed no significant change in discrimination. Even tests with unfiltered data indicate that the effect is present, although difficult to quantify, again due to saturation effects on the waveforms.  

As described by Plesset and Zwick \cite{Bib17} bubble growth starts with an initial, 
surface tension controlled stage, followed by an inertia controlled intermediate stage, 
which then leads to a heat transfer dominated final phase. The first two stages happen within the first 10 psec before the bubble reaches its critical radius and are probably not important here. During the third stage the bubble becomes macroscopic in size and its radius increases with time like $R(t) \propto t^{1/2}$. Within this model and given the thermodynamic parameters of C$_4$F$_{10}$ one can calculate growth rates, which exceed several micrometers per microsecond during the first 10 $\mu$s after nucleation. 
Since the acoustic pressure wave in the liquid is proportional to the acceleration with which the vapour phase grows \cite{Bib18}, it seems plausible that the very rapid bubble expansion contributes substantially to the acoustic signal during the first 50 $\mu$s and preserves a memory of the extension of the primary event with a resolution of tens of micrometers. The theory also predicts that the acceleration of the 
bubble wall is proportional to temperature, which is consistent with the observed increase in peak amplitude 
of the neutron and alpha induced signal.    
However a precise description of the relation between the produced pressure signal and the observed waveform requires more detailed studies of the transducer responses by using e.g calibrated hydrophones.     

\section{Conclusion\protect}
We observe a difference between the amplitudes of neutron and alpha particle induced events in superheated 
C$_4$F$_{10}$ in the sensitivity region interesting for dark matter searches. 
The discrimination effect was unexpected and was observed for the first time in a new generation of 
PICASSO modules with increased droplet sizes, such that all or a substantial fraction of the 
alpha tracks are contained.  The discrimination effect is not well understood yet, but could be explained by an 
increasing number of nucleation centres on the alpha track. With a better control of the acoustic coupling, improved 
long-term stability of the transducer system and reduced gain to avoid saturation effects, a better understanding and a further enhancement in discrimination can be expected. This effect has the potential to improve background suppression in future dark matter searches with superheated liquids, such as droplet detectors and bubble chambers. Other applications of the superheated liquid technique which could profit from this discrimination feature could be in the detection of traces of alpha-emitting actinides in biological samples \cite{Bib19, Bib20} and low level neutron counting. 

\begin{acknowledgments}
We wish to acknowledge the support of the Canadian National Science and Research Council (NSERC), 
the Canada Foundation for Innovation (CFI), the National Science Foundation (NSF), 
and the Czech Ministry of Education, Youth and Sports within the project MSM6840770029.
This work is also supported by NSF grant PHY-0555472 and the Indiana University South Bend Research and Development Committee.
We wish to thank Naomi Tankersley for technical support at IUSB.
\end{acknowledgments}


\begin{thebibliography}{12}

\bibitem{Bib1}
R. Apfel, Nucl. Inst. and Meth. 162 (1979) 603

\bibitem{Bib2}
H. Ing et al.; Radiat. Meas. 27(1997) 1

\bibitem{Bib3}
D.A. Glaser, Phys. Rev. 87 (1952) 665

\bibitem{Bib4}
V.Zacek, Il Nuovo Cimento 107A (1994) 1247

\bibitem{Bib5}
M. Barnab\'e-Heider et al; Phys.Lett.B624 (2005) 186

\bibitem{Bib6}
F. Seitz, The Physics of Fluids, Vol. 1, No. 1 (1958) 2


\bibitem{Bib7}
J. Ellis and R. Flores, Phys. Lett. B 263 (1991) 259


\bibitem{Bib8}
T.A. Girard et al ; Phys.Lett. B621 (2005) 233

\bibitem{Bib9}
E. Behnke et al; Science, Vol 319. (2008) 933 

\bibitem{Bib10}
M. Barnab\'e-Heider et al ; Nucl. Inst. and Meth. A555 (2005) 184

\bibitem{Bib11}
 J.E. Shepherd, Dynamics of vapour explosions: rapid evaporation and instability  of butane droplets at the superheat limit; PhD thesis, Caltech, 1980

\bibitem{Bib12}
R. Gornea, D\'etection directe de la mati\`ere sombre avec le d\'etecteur \`a gouttelette surchauff\'ees dans le cadre du projet PICASSO, PhD thesis, Universit\'e de Montr\'eal, 2008

\bibitem{Bib13}
K. Clark, A New and Improved Spin-Dependent Dark Matter Exclusion Limit Using the PICASSO Experiment; PhD thesis, Queen's University, 2008

\bibitem{Bib14}
D. Glaser et al; Phys.Rev. Vol. 102, No 6 (1956) 1653

\bibitem{Bib15}
A. Herve et al; Nucl. Instr. and Meth. 202 (1982) 417

\bibitem{Bib16}
E. Hugentobler, B. Hahn and F. Steinrisser, Helv. Phys. Acta 36 (1963) 601 

\bibitem{Bib17}
M.S. Plesset and S.A. Zwick, J. Appl. Phys. 23, 95 (1952)

\bibitem{Bib18}
L.D. Landau, E. M. Lifshitz, Fluid Mechanics, 2nd ed., Pergamon Press, Oxford, 
      1993, pp 251-283

\bibitem{Bib19}
C.K. Wang, W. Lim and L.K. Pan; Nucl. Inst. and Meth. A353 (1994) 524. 

\bibitem{Bib20}
D. Ponraju et al; Nucl. Inst. and Meth. A580 (2007) 388  



\end{thebibliography}

\end{document}